\documentclass[apj]{emulateapj}
\shortauthors{Allers et. al.}
\shorttitle{Near-IR Spectra of Young Brown Dwarfs}

\begin{document}

\title{Characterizing Young Brown Dwarfs using Low Resolution Near-IR Spectra}
\author{K.~N.~Allers} 
\affil{Institute for Astronomy, University of Hawaii, 
2680 Woodlawn Drive, Honolulu, HI 96822; allers@ifa.hawaii.edu}
\author{D.~T.~Jaffe}
\affil{Department of Astronomy, University of Texas at Austin, 
Austin, TX 78712-0259}
\author{K.~L.~Luhman}
\affil{Department of Astronomy and Astrophysics, The Pennsylvania State 
University, University Park, PA 16802}
\author{Michael~C.~Liu\altaffilmark{1,2}} 
\affil{Institute for Astronomy, University of Hawaii, 
2680 Woodlawn Drive, Honolulu, HI 96822}
\author{J.~C.~Wilson, M.~F.~Skrutskie, M.~Nelson, D.~E.~Peterson}
\affil{Department of Astronomy, University of Virginia, P.O. Box 3818, Charlottesville, VA 22903-0818}
\author{J.~D.~Smith, M.~C.~Cushing\altaffilmark{2}}
\affil{Steward Observatory, University of Arizona, Tucson, AZ 85721}

\altaffiltext{1}{Alfred P. Sloan Research Fellow}
\altaffiltext{2}{Visiting Astronomer at the Infrared Telescope Facility, which is operated by the University of Hawaii under Cooperative Agreement no. NCC 5-538 with the National Aeronautics and Space Administration, Office of Space Science, Planetary Astronomy Program.}

\begin{abstract}
We present near-infrared (1.0--2.4~$\mu$m) 
spectra confirming the youth and cool effective temperatures of 6 brown dwarfs and low mass stars 
with circumstellar disks 
toward the Chamaeleon II and Ophiuchus star forming regions.  
The spectrum of one 
of our objects indicates that it has a spectral type of $\sim$L1, making it 
one of the latest spectral type young brown dwarfs identified to date.  
Comparing spectra of young 
brown dwarfs, field dwarfs, and giant stars, 
we define a 1.49--1.56 $\mu$m H$_2$O index capable of determining 
spectral type to $\pm$1 
sub-type, independent of gravity.  We have also defined an index based on the 
1.14 $\mu$m sodium feature that is sensitive to gravity, but only weakly 
dependent on spectral type for field dwarfs.  Our 1.14 $\mu$m Na index can be 
used to distinguish young cluster members ($\tau \lesssim 5$~Myr) from 
young field dwarfs, both of which may have the 
triangular H-band continuum shape which persists for at least tens of Myr. 
Using T$_{eff}$'s determined 
from the spectral types of our objects along with luminosities derived from 
near and mid-infrared photometry, we place our objects on the H-R diagram and 
overlay evolutionary models to estimate the masses and ages of our young 
sources. 
Three of our sources have inferred ages ($\tau\simeq$10--30~Myr) 
significantly older 
than the median stellar age of their parent clouds (1--3~Myr).  For these three 
objects, we derive
masses $\sim$3 times greater than expected 
for 1--3~Myr old brown dwarfs with the 
bolometric luminosities of our sources.
The large discrepancies in the inferred masses and ages determined 
using two separate, yet 
reasonable methods, emphasize
the need for caution when deriving or exploiting brown dwarf mass and age estimates. 

\end{abstract}


\section{Introduction}
In recent years, surveys have begun to 
find young brown dwarfs with spectral types extending into the 
L dwarf regime \citep{kirkpatrick06,luhman06,lucas01}.  
The low gravity, and hence youth, of these objects has typically 
been determined via a triangular 
H-band continuum shape which exists for brown dwarfs until they are at least 
tens of Myr old \citep[log $g \lesssim$ 4.5][]{kirkpatrick06}.
The masses of these sources have been estimated 
by placing them on an H-R diagram, with temperatures 
determined by either fitting theoretical spectra to observed spectra 
\citep{lucas01} or by using a spectral~type-temperature 
relationship that is intermediate between dwarf and giant stars 
\citep{luhman06}.  
Both of these methods have systematic uncertainties which grow as one moves to 
later spectral types.

Spectral types of several young, late-M type brown dwarfs 
have been determined by comparing observed optical spectra to spectra of 
giants and dwarfs, with the best fit usually being an average of giant and 
dwarf spectra \citep{guieu06,luhman04c,briceno02}.  
The spectral energy distributions 
(SED's) of late-M type objects peak in the near-IR 
\citep{cushing05}, making them, particularly in the presence of 
foreground reddening, easier to observe in the near-IR than in the optical.  
An infrared spectral type classification scheme for young, 
late-M type objects has been developed by 
comparing their infrared spectra to those of 
optically classified young objects \citep{luhman04a}.  
Spectral types can then be converted to 
T$_{eff}$ using a temperature scale \citep[e.g.][]{sluhman03} which 
is intermediate to SpT-T$_{eff}$ relationships for field 
M dwarfs \citep{golimowski04} and M giants \citep{perrin98}.  

The tools available for classifying young L type brown dwarfs 
are considerably less well-developed than the tools for classifying young 
M stars and brown dwarfs.   The L subclass for field dwarfs 
is defined by the disappearance of TiO and VO bands as seen in their optical 
spectra \citep{kirkpatrick99}.
To date, however, there are only five young, optically classified L dwarfs, 
three of which are field objects displaying spectral signatures of youth 
\citep{kirkpatrick06,luhman06,kirkpatrick01} and two of which were 
recently identified 
in nearby star-forming regions \citep{allers06, jayawardhana06}.  
The young L dwarfs identified to date are early-L type (L0-L2); 
it remains unclear how to classify cooler young brown dwarfs, 
if and when they are found.
There are no L giant stars that could be
combined with field dwarf observations to fit a spectral type to young 
objects.  Even if a spectral type for a young L dwarf could be determined, 
the SpT--T$_{eff}$ relationship is unknown for young L brown dwarfs, 
and there is no L-giant sequence to which one could compare the 
temperature--spectral type relation.
In this paper we discuss the spectral properties of a sample of six M6--L1 
young brown dwarfs and low mass stars identified by \citet{allers06}, 
and by comparison to field dwarfs and giants, develop an empirical 
scheme for determining the SpT and verifying the low-gravity of young objects, 
with a view toward studying young objects of even later spectral type.  
We derive effective temperatures using the \citet{sluhman03} 
SpT-T$_{eff}$ relation and 
place our objects on an H--R diagram.  We then critically examine the 
disagreements 
for a few objects between the masses and ages derived from the H--R diagram and those 
reported in \citet{allers06}.

\section{A Sample of Young Brown Dwarfs}
The Spitzer Legacy Program, ``From Molecular Cores to Planet Forming Disks'' 
\citep[c2d;][]{evans03} provides IRAC and MIPS imaging of 
5 different star forming regions
covering a total of $\sim$16 square degrees.
\citet{allers06} conducted a deep near-IR survey (10 $\sigma$ limits of 
better than 
22.9, 19.7, 19.1 and 18.6 magnitudes in I, J, H, and Ks respectively),
covering 1.5 square degrees towards 
regions in the Chamaeleon II, Ophiuchus, and Lupus I molecular clouds lying 
within the Legacy Program boundaries, having 
modest total extinction \citep[A$_V<$ 7.5 magnitudes;][]{cambresy99} and 
containing or close to known
T~Tauri stars \citep{wilking87, hartigan93, schwartz77}.   
The survey resulted in a sample of 19 candidate 
young stars and brown dwarfs with 
disks \citep{allers06} found by dereddening objects to the I, J, H and Ks 
colors of young brown dwarfs, 
and selecting objects showing mid-IR excess emission.  
The mid-IR excess assured that the objects were young, assuming that the 
disk lifetimes are comparable to those of low-mass YSOs \citep[$\lesssim$6~Myr;][]{haisch01}.
The 
luminosities of objects in the sample were determined 
by integrating the dereddened 
source flux from 0.8 
to 3.6 $\mu$m. Average uncertainties in luminosity were $\pm$0.2 dex in 
log(L$_{\ast}$).  The T$_{eff}$, mass 
and log~$g$ of the objects were estimated from the luminosity using 
model isochrones \citep{baraffe03} and assuming an age of 3~Myr for 
Chamaeleon II and 1~Myr for Ophiuchus and Lupus I.  
The faintest source in the \citet{allers06} survey showing evidence of a 
circumstellar disk from 5.8 to 24 $\mu$m 
has an observed Ks magnitude of 15.98.  All of the sources 
in the 
sample are well above their detection limits in the near-IR, 
thus uncertainties in
their photometry are dominated by systematics, approximately 0.05, 
0.03, 0.03, \& 0.03 magnitudes in I, J, H, and Ks respectively.  
The \citet{allers06} sample 
has luminosities of 0.56~$>$~log(L$_{\ast}$/L$_{\odot}$)~$>$~-3.11 
and five sources with possible masses 
below 12 M$_J$.

\section{Near-IR Spectra}
\subsection{Observations and Data Reduction}
We obtained near-IR (0.9--2.5~$\mu$m) 
spectra of 6 candidate young brown dwarfs in 
the Ophiuchus and Chamaeleon II molecular clouds 
(\#1, \#2, \#5, \#11, \& \#14 in \citet{allers06} along with one 
additional object 1.9\arcsec\ north of \#5).  Our spectra were taken 
on the nights of 2005 April 29 through May 1 
using the CorMASS spectrometer\footnote{CorMASS 
is supported by a generous gift to the University of Virginia
Astronomy Department from the F.H. Levinson Fund of the Peninsula
Community Foundation.} \citep{wilson01} on the Landon Clay (Magellan II) 
telescope at Las Campanas Observatory.  
For comparison, we also obtained CorMASS spectra of 4 young 
brown dwarfs in Chamaeleon I with spectral types known from optical 
observations \citep[Cha H$\alpha$ 10 (M6.25), 
Cha H$\alpha$ 11 (M7.25), CHSM 17173 (M8), OTS44 (M9.5);][]{luhman04c, 
luhman04a}, 3 field dwarfs \citep[Gl 406 (M6V), 
LHS 2065 (M9V), Kelu-1 (L2V);][]{leggett01,leggett02}, and a late-type giant 
\citep[Br1219-1336 (M9III);][]{kirkpatrick97}.  The details of our 
observations are listed in Table \ref{sobservations}.  For most targets, 
two exposures were obtained with the object on the slit followed by 
one exposure with the object off the slit, and this pattern was 
repeated as necessary. The spectra were reduced with a modified version of 
the Spextool package \citep{cushing04}, which included a correction for 
telluric absorption following the method described in \citet{vacca03}.  

\begin{deluxetable}{lccl}
\tablecolumns{4}
\tabletypesize{\footnotesize}
\tablecaption{CorMASS Observations}
\tablewidth{0pt}
\tablehead{
\colhead{Source} &
\colhead{Date}             &
\colhead{Airmass}             &
\colhead{N $\times$ exptime}
}
\startdata
cha1257-7701   &    042905 & 1.49 & 6x180\\
cha1258-7709   &    050105 & 1.52 & 4x45\\
cha1305-7739   &    042905 & 1.51 & 6x180\\
oph1622-2405B  &    042905 & 1.01 & 4x120\\
oph1622-2405A  &    042905 & 1.01 & 2x180,2x120\\
oph1623-2338   &    042905 & 1.03 & 4x120\\
\\
Cha H$\alpha$ 10    &    042905 & 1.53 & 6x120\\
Cha H$\alpha$ 11    &    042905 & 1.52 & 6x120\\
CHSM 17173     &    042905 & 1.48 & 6x120\\
OTS44       &    043005 & 1.49 & 6x120\\
\\
Gl 406       &    050105 & 1.26 & 4x0.5\\
LHS 2065     &    050105 & 1.13 & 4x20\\
Kelu--1       &    043005 & 1.01 & 4x90\\
\\
BR 1219-1336 &    050105 & 1.06 & 4x1
\enddata
\label{sobservations}
\end{deluxetable}

\begin{figure} 
\epsscale{1.0}
\plotone{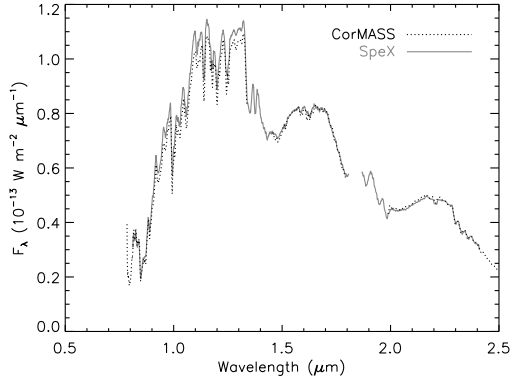}
\caption[Comparison of CorMASS SpeX spectra]
{\label{spex} Comparison of flux calibrated spectra from 
CorMASS (black-dotted line) and SpeX \citep[gray line;][]{cushing05} spectra.}
\end{figure}

We establish the absolute flux calibration of our R$\simeq$300 CorMASS
spectra in W m$^{-2}$ $\mu$m$^{-1}$ by calculating 
$\langle$F$_{\lambda}$$\rangle$ in J, H, and Ks 
from photometry of our sources \citep{allers06}.
For each spectrum, 
we use 2MASS photometry of young, Chamaeleon I 
standards or our own near-IR photometry \citep{allers06} to calculate the 
flux calibration in J, H, \& Ks and we apply the average correction of the 
three bands to the entire near-IR spectrum.
A comparison of a calibrated CorMASS spectrum to a SpeX spectrum 
of LHS 2065 \citep{cushing05} is shown in 
Figure \ref{spex}. 
We used near-IR photometry of LHS 2065 from \citet{leggett01} to calibrate our 
CorMASS spectrum into absolute flux units. The SpeX spectrum of LHS~2065 
was flux-calibrated using an observed A0V star 
\citep{cushing05}, and has 
only been smoothed to R$\simeq$300---we did not alter the flux levels.
There is good agreement between 
not only the flux levels of the spectra, but also the spectral 
shapes and features.  The LHS~2065 CorMASS flux is slightly ($\sim$5\%) 
below the SpeX flux 
in the J band.  This level of difference 
is not surprising given that the CorMASS observations were not 
taken with the slit aligned to the parallactic angle. 
To quantify the importance of the spectral tilt added by differential 
atmospheric refraction and/or poor instrumental 
flux calibration, we can deredden the spectra to achieve a best match 
with the SpeX results for field dwarfs where spectra taken with 
both instruments are available, and use the A$_V$ correction as a measure of 
the uncertainty due to spectral tilt.
For LHS~2065, the H and K band fluxes in the CorMASS and SpeX spectra agree to 
within 1\%; 
dereddening its CorMASS spectrum by A$_V$~=~0.2~magnitudes 
(using the reddening law of \citet{fitzpatrick}), increases its J-band flux 
so that it agrees with its SpeX spectrum.  Similar comparisons of CorMASS and 
SpeX spectra for Gl 240 (M6V) and Kelu-1 (L2V) \citep{leggett02} 
show that the absolute flux calibrations 
of SpeX and CorMASS agree to within 10\%, and any differences in the 
shape of the spectra can be resolved by reddening or dereddening the 
CorMASS spectra by A$_V < 0.3$.

We also obtained observations of 
G196-3B, a wide L dwarf companion to the young 
($\tau=\sim$100~Myr) M dwarf G196-3 \citep{rebolo98}.  Our near-IR spectrum of 
the object was taken on 2002 May 30 UT using the SpeX spectrograph 
\citep{rayner03} on NASA's 
Infrared Telescope Facility (IRTF), located on the summit of Mauna Kea, Hawaii.
We used the SXD mode to obtain simultaneous spectra from 1.1--2.4~$\mu$m.
The slit was aligned with the parallactic angle, and the target was 
observed at an airmass of 1.3--1.4.  We used the 0.3\arcsec\ wide slit, 
producing a spectum with a resolution ($\lambda/\Delta\lambda$) of $\sim$2000.  
We obtained nine two-minute exposures for a total integration time of 
18~minutes, with the object dithered along the slit between exposures.  
We observed a nearby A0 star for telluric calibration.  The spectra were 
processed using the facility reduction pipline Spextool \citep{cushing04}.

\subsection{Spectral Types and Extinction}
\begin{figure}
\plotone{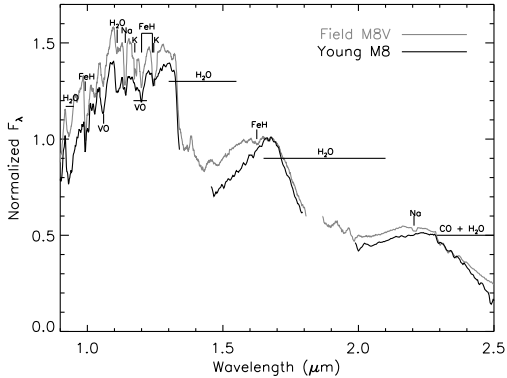}
\caption[Comparison of an M8 Field Dwarf and a Young M8 object]
{\label{gravity} Comparison of a young brown dwarf and a higher gravity 
field dwarf at the same optical spectral type.  
The black line shows the CorMASS spectrum of  
CHSM 17173, a young M8-type brown 
dwarf in Chamaeleon I \citep{luhman04c}.
The gray line shows the SpeX spectra of vB10, an M8V type field 
dwarf \citep{cushing05}.}
\end{figure}

Even at low-resolution, there are significant 
differences between the infrared spectra of young brown dwarfs 
and older field dwarfs.  Figure \ref{gravity} shows the CorMASS spectrum 
of CHSM 17173, a young M8 object in Chamaeleon I \citep{luhman04c}, along 
with the SpeX spectrum (smoothed to R$\simeq$300) of the M8V field dwarf, vB10 
\citep{cushing05}.  
Certainly the most dramatic difference between the two spectra is 
in the H band.  The young brown dwarf has a much more sharply peaked H band 
spectrum, whereas the field dwarf has a plateau from $\sim$1.6--1.7~$\mu$m.
While the spectra drop equally quickly at the long wavelength end of the H 
band, the short wavelength end of this band has much deeper H$_2$O absorption 
in the spectrum of the young brown dwarf.
The triangular shape of the H band for young brown dwarfs has been noted in 
previous studies \citep[e.g.][]{luhman04a, lucas01}. 
The K band spectrum of the young brown dwarf reaches a maximum at a 
longer wavelength than the field dwarf spectrum, and 
the young M8 spectrum drops off more steeply from 2.3 to 2.5 $\mu$m than 
the field dwarf spectrum.  
The CHSM 17173 spectrum 
does not have any trace of the 2.2~$\mu$m NaI absorption feature which is 
still readily apparent in the spectrum of vB10.  
The strongest features in the J band spectra of late M field 
dwarfs include absorption bands of H$_2$O (0.92 and 1.11 $\mu$m), 
VO (1.07 and 1.18 $\mu$m), and FeH (0.99, 1.20, \& 1.24 $\mu$m) as well as 
neutral metal lines of NaI (1.14 $\mu$m) and KI (1.17 and 1.25 $\mu$m).
The young brown dwarf shows deeper and wider 
H$_2$O features than the field dwarf.  
The VO absorption is deeper at 1.07 $\mu$m in the young brown dwarf than the 
field dwarf, and is blended with KI and FeH at 1.18 $\mu$m.  
The young brown dwarf does not have as deep of NaI features as the 
field dwarf, and shows no noticeable 1.17 $\mu$m KI absorption.  
The 1.25 KI/FeH and 1.18 
KI/FeH/VO blended 
features are much more prominent in the field dwarf than in the young 
object.  
The noticable differences between the young M8 brown dwarf 
and field M8 dwarf spectra indicate that we can use the R$\simeq$300 spectra 
to confirm the youth of our sample, independent of the presence of 
circumstellar disks.

\begin{figure}
\plotone{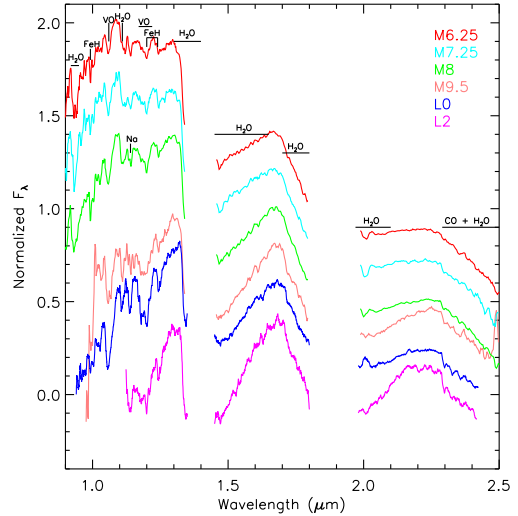}
\caption[Young Object Spectral Types]
{\label{young} Our young standards with optical spectral types: Cha H$\alpha$ 10 (M6.25), Cha H$\alpha$ 11 (M7.25), CHSM 17173 (M8) \citep{luhman04c}, OTS44 (M9.5; Luhman, {\it in preparation}), 2MASS0141--4633 \citep[L0;][]{kirkpatrick06} and G196--3B \citep[L2;][]{rebolo98,kirkpatrick01}}
\end{figure}

We determine the spectral types of our objects by comparing their 
near-IR spectra to spectra of young brown dwarfs in Chamaeleon I with known 
spectral types determined from optical spectra 
\citep[Cha H$\alpha$ 10 (M6.25), 
Cha H$\alpha$ 11 (M7.25), CHSM 17173 (M8);][OTS44 (M9.5); 
Luhman, {\it in preparation}]{luhman04c}.
At $\sim$2~Myr old \citep{luhman04c}, the Chamaeleon I cloud is roughly the 
same age as Ophiuchus ($\sim$1~Myr) and Chamaeleon II ($\sim$3~Myr). Thus, 
our Chamaeleon I standards likely have gravities similiar to the 
sources in our sample, making them acceptable comparison sources.
The spectral types of Cha H$\alpha$ 10, Cha H$\alpha$ 11, CHSM 17173, 
and OTS44 were determined by comparison of their low resolution 
red optical spectra to the average spectra of field 
dwarfs and giants at a given spectral type \citep[][Luhman et al. in prep.]{luhman04c}.
\citet{comeron00} determine later spectral types for 
Cha H$\alpha$ 10 and Cha H$\alpha$ 11 (M7.5 and M8 respectively) than those 
published by \citet{luhman04c}, possibly due to their
comparison to the optical spectra of field dwarfs only.
To determine spectral types later than M9.5, we also include SpeX 
near-IR spectra (smoothed to R$\simeq$300) 
of two young field brown dwarfs, 2MASS0141--4633 
\citep[L0;][]{kirkpatrick06} and G196--3B \citep[L2;][]{rebolo98,kirkpatrick01}.
Though 2MASS0141--4633 and G196--3B are older
\citep[1--50~Myr and 60--300~Myr old;][]{kirkpatrick06,kirkpatrick01,rebolo98} 
than our objects in Chamaeleon II and Ophiuchus, their spectra show 
the distinct triangular H-band shape indicative of youth.
The CorMASS spectra of Cha H$\alpha$ 10, 
Cha H$\alpha$ 11, CHSM 17173, and OTS44, dereddened by 
their published A$_V$'s \citep{luhman04c,luhman04a}, along with SpeX spectra of 
2MASS0141--4633 and G196--3B smoothed to R$\simeq$300, are displayed in 
Figure \ref{young}.

\begin{figure*}
\plotone{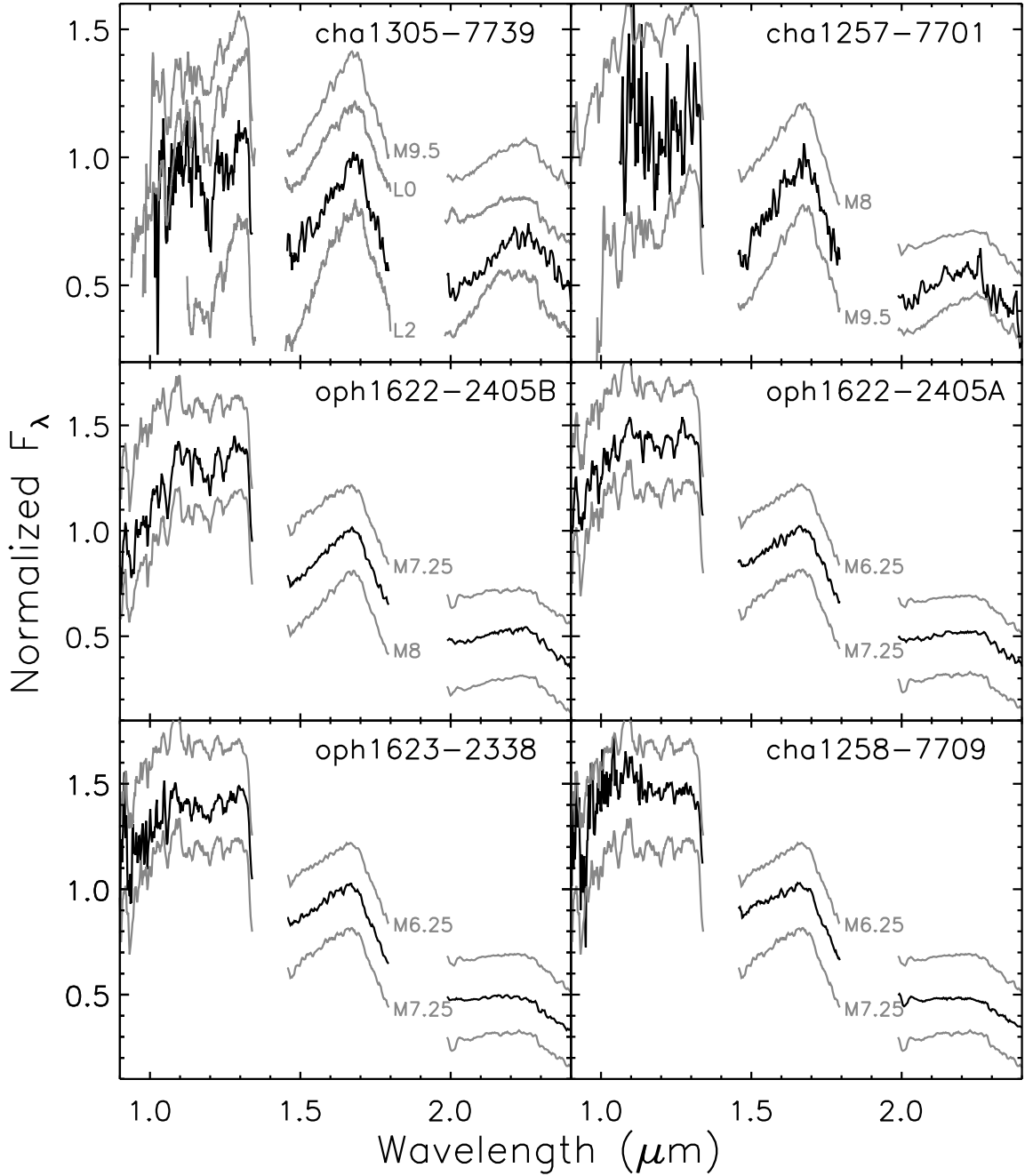}
\caption[Six New Young Brown Dwarfs]
{\label{spT} Dereddened near-IR spectra of our candidate young brown
dwarfs (black lines), compared to spectra of young
brown dwarf standards shown in Figure \ref{young} (gray lines).
The values of A$_V$ by which we dereddened the spectra along with our
adopted spectral types are listed in Table \ref{teff}.}
\end{figure*}

We determine the reddening of the objects in our sample by comparison to each of the young standards using the reddening law of \citet{fitzpatrick}, starting with the 
A$_V$ from \citet{allers06} and adjusting the value of A$_V$ until the 
fluxes (normalized at 1.68~$\mu$m) in the J and H bands agree with the 
fluxes of the standard.  
Our method assumes that the circumstellar disk makes a negligible contribution 
to our near-IR spectra, a reasonable assumption given that brown dwarfs disks 
are too cool to provide any significant near-IR excess emission
\citep{liu03,natta01}, particularly in the J and H bandpasses.  
The continuum shape of the spectra are very sensitive to A$_V$; we can 
easily distinguish between A$_V\pm0.5$.
After dereddening our spectra, we compare 
the overall shape of the spectra and their molecular features.  
The shape of the H band changes substantially from M6.25 to L2, as the 1.4
to 1.65 $\mu$m H$_2$O features become deeper and create 
a triangular shape to the band.   
The H$_2$O absorption on the short wavelength side of the K band also changes 
with spectral type.  The CO bandhead at 2.3 $\mu$m becomes 
more noticeable at later spectral types, having a sharp difference from M6.25 
to M7.25.  In the J band, the FeH and VO features become more prominent with 
absorption features at 1.07, 1.20 and 1.24~$\mu$m becoming wider and deeper at 
later spectral types.  Looking at all of these features together, we can 
distinguish SpT to within $\pm$1 subclass.  For sources with spectral types 
later than M9.5, we are dependent on comparisons to young field objects.  
Older field dwarfs have intrinsically bluer near-IR spectra and shallower 
H$_2$O absorption features than younger brown dwarfs, effects noted by \citet{kirkpatrick06} and \citet{mcgovern04}.  Thus, though they have the triangular H-band shape indicative of youth, the spectra of our 
young field standards are likely 
intrinsically bluer and have shallower H$_2$O absorption features than 
would be expected for younger ($\lesssim$5~Myr) cluster brown dwarfs of 
the same spectral type.
Figure \ref{spT} displays our dereddened spectra, along with 
the dereddened spectra of the closest-matching young brown dwarf standards 
(Figure \ref{young}).  Spectral types and A$_V$'s 
for our objects are listed in Table \ref{teff}.  A discussion of our 
derivation of spectral type for each object 
(ordered from earlier to later spectral type) follows.

\subsubsection{Cha1258-7709}
Cha1258-7709 was first identified as the young brown dwarf candidate, ISO-ChaII~13, based on its large mid-IR excess \citep{persi03}.
We deredden cha1258-7709 by A$_V=10.4$ to make the J and H band flux levels 
(normalized to F$_{\lambda}$ at 1.68$\mu$m) agree with the flux levels 
of Cha~H$\alpha$~10.  
The slope of the H and K band H$_2$O features in cha1258-7709 are 
best matched to the spectrum of Cha~H$\alpha$~10 (M6.25), but are slightly 
less steep.  The triangular shape of the H band spectrum confirms this 
object's youth.  
The CO bandhead at 2.3 $\mu$m is slightly 
more pronounced in the spectrum of cha1258-7709 than for Cha~H$\alpha$~10. 
The J-band FeH and H$_2$O features are well matched.  Overall, the 
cha1258-7709 spectrum matches the spectrum of Cha~H$\alpha$~10 (M6.25) 
quite well, but since the H and K band slopes are slightly shallower in 
cha1258-7709, we assign a spectral type of M6. Our spectral type determination 
is slightly earlier than that reported from its optical spectrum 
\citep[M7;][]{alcala06}, though the 
two spectral types agree to within our uncertainties.  
\citet{alcala06} find a best fit A$_V$ of 5.0$\pm$0.5 for cha1258-7709, while 
we determine an A$_V$ that is more than a factor of 2 larger, and in 
better agreement with the photometrically determined A$_V$ of 
\citet{allers06}.  Dereddening the near-IR spectrum of cha1258-7709 by an 
A$_V$ of 5.0 rather than 10.4 does not allow a reasonable fit to any of 
our near-IR young spectral standards.  

\subsubsection{Oph1623-2338}
The spectrum of oph1623-2338, dereddened by A$_V=8.2$ has its closest match 
to the spectrum of Cha~H$\alpha$~11 (M7.25).  
The slopes of the H$_2$O features 
in the H and K band are much shallower than for a young M8 object, and are 
somewhat shallower 
than Cha~H$\alpha$~11.  Comparison to a young M6.25 spectrum, however, yields 
a poor match of the FeH features and the 2.3 $\mu$m CO bandhead, 
which are much deeper for 
oph1623-2338.  The CO bandhead agrees best with the spectrum of 
Cha~H$\alpha$~11 (M7.25).  We assign a spectral type of M7 to oph1623-2338.

\subsubsection{Oph1622-2405A}
Oph1622-2405A is not in the \citet{allers06} sample because in the c2d 
catalogs, its 5.8 and 
8.0~$\mu$m mid-IR fluxes were not resolved from a source 1.9\arcsec\ to 
the south. The binary nature of oph1622-2405 is discussed in \citet{allers05}, \citet{binary}, \citet{close07}, and \citet{jayawardhana06b}.
The spectrum of oph1622-2405A, dereddened by A$_V$=-0.2, 
is best matched to Cha~H$\alpha$~11 (M7.25).  The published A$_V$ value
for Cha~H$\alpha$~11 \citep[A$_V = 0.3$;][]{luhman04c} might be 
underestimated, and an A$_V$ of -0.2 for an unreddened object 
is well within our uncertainty of A$_V\pm$0.5.  We assign an A$_V$ of 0.0 for 
oph1622-2405A.
The H$_2$O, VO, and CO features of oph1622-2405A agree 
quite well with Cha~H$\alpha$~11.
We assign a spectral type of M7 to oph1622-2405A.  
Our spectral type is in agreement with the spectral type derived from its optical spectrum (M7.25) by \citet{binary}, but is earlier than the spectral type 
(M9) reported by \citet{jayawardhana06b}.  
The 1.14 $\mu$m NaI feature is 
slightly stronger and the 1.11~$\mu$m H$_2$O feature is slightly weaker 
in oph1622-2405A than in Cha~H$\alpha$~11, which may 
be indicative of a slightly higher gravity (though still well below the 
gravities of field dwarfs) for oph1622-2405A.

\subsubsection{Oph1622-2405B}
Oph1622-2405B dereddened by A$_V=1$ matches the spectrum of CHSM 17173 (M8) 
almost perfectly.  The band shapes, FeH, VO, and CO features 
all agree very well between oph1622-2405B and CHSM 17173, so we assign a 
spectral type of M8 to oph1622-2405B.  As is the case for oph1622-2405A, the 
J-band NaI and H$_2$O features in oph1622-2405B hint at a slightly 
higher gravity than CHSM 17173.  Based on optical follow-up 
this souce has a spectral type of M8.75-M9 
\citep{binary,jayawardhana06}, which agrees with our spectral type determination
to within the uncertainties.

\subsubsection{Cha1257-7701}
The spectrum of cha1257-7701 has the lowest S/N of any of our CorMASS spectra, 
so comparison of the FeH and CO features is difficult, and we must base 
our comparison on the continuum shapes of the bands.  
Compared to OTS44 (M9.5) and 
2MASS0141--4633 (L0), cha1257-7701 
dereddened by A$_V=3.0$ has similar H and K band shapes, but its flux level 
from 1.0 to 1.2 $\mu$m is noticibly higher than for both OTS44 and 
2MASS0141--4633, which indicates 
an earlier spectral type (notice the J-band spectral shapes in Figure 
\ref{young}).  The J-band continuum shape of cha1257-7701, dereddened by 
A$_V$=5.5, matches CHSM~17173 fairly well, but the H and K band H$_2$O 
absorption features are noticibly deeper in cha1257-7701.
The spectrum of cha1257-7701 appears intermediate to that of 
CHSM~17173 (M8) and OTS44/2MASS0141--4633 (M9.5/L0).
We thus assign a spectral type of M9 and an A$_V$ of 4.0$\pm$1.0 to cha1257-7701.  
Our spectral type 
agrees with the spectral type of cha1257-7701 determined from its optical 
spectrum \citep{jayawardhana06}.

\subsubsection{Cha1305-7739}
The H-band spectrum of cha1305-7739 (dereddened by A$_V=3$) 
is fairly similar to that of OTS44 (M9.5).  The 1.5--1.7~$\mu$m slope of cha1305-7739 is 
steeper than 2MASS0141--4633 (L0), but clearly not as steep as G196--3B (L2).  
Cha1305-7739 has a more sharply peaked K band spectrum than both OTS44 and 
2MASS0141--4633, 
and deeper and wider 1.2 $\mu$m VO/FeH/KI and 1.25 $\mu$m FeH/KI 
features, 
indicating a later spectral type for cha1305-7739 than OTS44 and 
2MASS0141--4633.  In general, the spectrum of cha1305-7739 appears 
intermediate between 2MASS0141--4633 and G196--3B.  We assign a 
spectral type of L1 to cha1305-7739, making it among the latest 
type young brown dwarfs discovered to date.  
The strengths of the H$_2$O absorption features in the near-IR are 
known to become weaker for higher gravity objects \citep[e.g.][]{mcgovern04}.  
Our young field standards are likely to have slightly higher gravities, 
and thus shallower H$_2$O absorption features at a given spectral type, 
than young objects in Chamaeleon II and Ophiuchus.  As a result, using 
the young field dwarfs as standards for determining 
the spectral type of cha1305-7739 could result in a spectral type determination 
that is too late, though the magnitude of this effect is unknown.
The L spectral class for field dwarfs is defined by the 
absence of TiO in their optical spectra.  
The spectral type determined from the optical spectroscopy followup of cha1305-7739 is L0$\pm$2 \citep{jayawardhana06}; in good agreement with our 
near-IR determined spectral type.

\subsubsection{Spectral Type Sensitive Indices}
\begin{figure}
\plotone{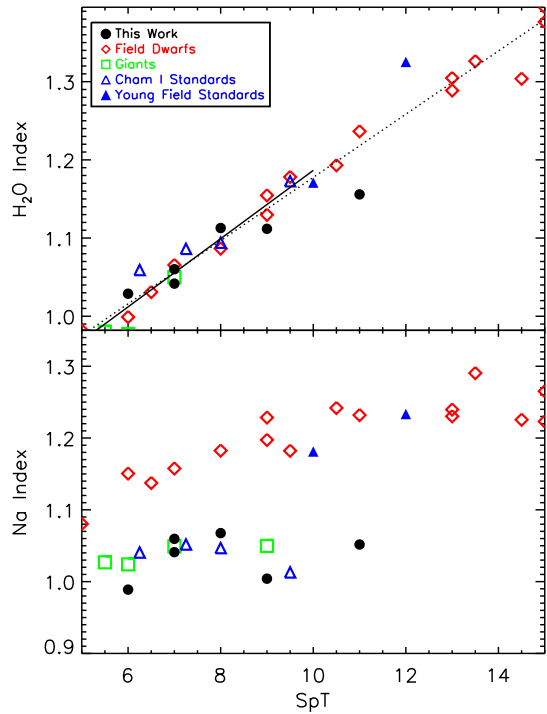}
\caption[Spectral Type and Gravity Sensitive Indices]
{\label{indices} SpT and gravity sensitive indices.
The spectral types range from M5 (SpT=5) to L5 (SpT=15).
Our sources, for which spectral types are determined from CorMASS
spectra, are shown as solid black circles.  The red diamonds
are values for field dwarfs from \citet{cushing05}, the green squares are for
giants, and the blue triangles are for young standards in the Chamaeleon I star forming region (open triangles) and for young field dwarf standards (filled triangles). Top: 
The H$_2$O index, as defined in this work, is
$\langle F_{\lambda=1.550-1.560}\rangle/\langle F_{\lambda=1.492-1.502}\rangle$.
The solid line shows a least-squares linear fit to the
field dwarfs, giants and young
standards for spectral types M5--L0.  The dotted line shows the fit to
field dwarfs from M5--L5.  Bottom: The Na index is
$\langle F_{\lambda=1.150-1.160}\rangle/\langle F_{\lambda=1.134-1.144}\rangle$.
}
\end{figure}

Several near-IR spectral indices have been suggested for use in 
determining spectral type.  The H$_2$O bands are commonly used in 
obtaining near-IR 
spectral indices used for classification, particularly for field 
brown dwarfs \citep{geballe01, reid01, delfosse99}.  We tested several 
suggested H$_2$O indices found in the literature 
\citep{gorlova03, geballe01, reid01, lucas01, delfosse99} on 
our spectra of field dwarfs, young dwarfs and giants, and found that the
index-SpT relationships based on these indices are highly gravity dependent. 
Thus, determining the spectral type from H$_2$O indices in the literature 
requires prior knowledge about the gravity of the object.  Given the 
differences in the shape of H$_2$O absorption features between field and 
young dwarfs (Figure \ref{gravity}), this is not surprising.  One feature to 
note in Figure \ref{gravity} is that the slope of the blue end of the H band 
($\sim$1.50--1.57 $\mu$m) is very similar in the two spectra,
hinting that an H$_2$O index measuring the 
slope of this feature might be insensitive to gravity.
By combining our 
data for field dwarfs, young brown dwarfs and giants, along with additional 
field dwarf and giant spectra from the IRTF Spectral Library 
\citep{cushing05}, we can test indices as a function of spectral type and 
gravity.   
We find that a spectral index defined as $\langle F_{\lambda=1.550-1.560}\rangle/\langle F_{\lambda=1.492-1.502}\rangle$ yields
an index-SpT relationship that is independent of gravity 
(Figure \ref{indices}).  
Fitting a line to the index versus spectral types of 
field dwarfs, giants, and young standards for spectral types of M5--L0 
(solid line in Figure \ref{indices}), we find a SpT--index relation: 
\begin{equation}
\frac{\langle F_{\lambda=1.550-1.560}\rangle}{\langle F_{\lambda=1.492-1.502}\rangle} = 0.75(\pm0.03) + 0.044(\pm0.004)\times SpT
\end{equation}
Our fit includes uncertainties in SpT of $\pm$0.5 for field dwarfs and 
giants (John Rayner and J. Davy Kirkpatrick, personal comm.), $\pm$0.25 
for our Chamaeleon I standards \citep{luhman04c}, and $\pm$1 for 
2MASS0141--4633 (J. Davy Kirkpatrick, personal comm.), as well as 
uncertainties in the index calculated from the spectra themselves.
The dispersion of the field dwarfs, giants, and young standards 
about the linear fit is $\pm$ 0.017 in the index.
For an object with a spectral type of M8, the uncertainties in the 
linear fit correspond to uncertainties of $\pm$0.04
in the index and $\pm$1.0 in the sub-type.  The S/N per resolution element 
(at R$\simeq$300) at the wavelength of the feature needed to achieve 
an uncertainty in the spectrum equal the 
uncertainty in the fit to the data is $\sim$30.  Changing the A$_V$ by 
which the spectra are dereddened by $\pm$1 magnitude changes the H$_2$O 
index by 0.011.
We also show a linear fit 
to the field dwarf indices for SpT's of M5--L5 
(dotted line in Figure 
\ref{indices}), giving a SpT--index relation for M5--L5 field dwarfs of:
\begin{equation}
\frac{\langle F_{\lambda=1.550-1.560}\rangle}{\langle F_{\lambda=1.492-1.502}\rangle} = 0.77(\pm0.02) + 0.040(\pm0.002)\times SpT
\end{equation}
The SpT--index relations for M5--L5 field dwarfs and M5--L0 field dwarfs, 
giants, and young standards are very similar. This hints that 
our SpT--index relation could be used to derive the spectral types of 
young brown dwarfs later than L0 (for which young, optically-classified 
standards in clusters do not yet exist).
The SpT of our sources 
derived from the two relations differ by 0.25 sub-types at most, which is 
much less than the uncertainties from the linear fit.  The spectral types 
of the sources in our sample, as determined from the value of the index, 
agree very well with the spectral types determined by matching their entire 
near-IR spectra to our young standards.  The largest deviation is 1.5 subtypes,
but the others deviate by less than 0.7 subtypes, well within our $\pm$1 subtype uncertainty in spectral type determination.  

\subsubsection{Gravity sensitive indices}

Differences in the gravities of young brown dwarfs, field dwarfs and giants 
are evident in several near-IR spectral features.  
Alkali metal lines of KI and NaI 
\citep[e.g.][]{mcgovern04,gorlova03, kirkpatrick06, lucas01} 
are stronger in objects 
with higher gravity, as a larger amount of KII and NaII can recombine at the 
higher atmospheric pressures in field dwarfs to form neutrals. 
 \citet{gorlova03} derive a relationship for log~$g$ with the 
pseudo-equivalent width of KI, but limit it to spectral types between M6 and M8.
FeH lines are stronger in 
high-gravity objects, but also vary with spectral type 
\citep{mcgovern04,gorlova03}.
Other possible gravity indicators are VO and TiO \citep{mcgovern04}, 
which are weaker at higher gravities due to condensation effects.
The 2.3 $\mu$m CO bandhead has long been used as a gravity indicator in stellar 
objects \citep[e.g.][]{kleinmann86}.
We tested indices of KI, NaI, VO, FeH, and CO, and found that the only index 
which had little correlation with SpT and had 
well separated index values for high (field dwarf) and low (young objects and 
giants) gravity objects was one based on the 1.14~$\mu$m NaI feature.
We chose to use indices of flux ratios rather than equivalent widths, since 
the continuum level in R$\simeq$300 near-IR spectra is masked by the wealth of 
atomic and molecular absorption features.
The bottom panel 
of Figure \ref{indices} shows our gravity sensitive index, 
defined by the depth of the 1.14 $\mu$m NaI feature,
$\langle F_{\lambda=1.150-1.160}\rangle/\langle F_{\lambda=1.134-1.144}\rangle$. Field brown dwarfs have an Na index with values $\sim$1.15 at M6, increasing gradually to $\sim$1.28 at L5.  The giants and young brown dwarfs have an Na index $\sim$1.03 over the range from M8 to L1. The gravity sensitivity of NaI has been noted previously using the 
2.2~$\mu$m line \citep{gorlova03}.  The K band for young objects could 
suffer from veiling, so we advocate the use of the 1.14 $\mu$m NaI line to 
verify the low gravity of sources.

Interestingly, the young field dwarfs 
2MASS0141--4633 and G196--3B have NaI indices that agree with older, 
higher-gravity field dwarfs, despite having the triangular H-band shape 
indicative of youth.  Given that 2MASS0141--4633 and G196--3B are $\sim$1--50
and $\sim$60--300~Myr old \citep{kirkpatrick06,kirkpatrick01,rebolo98}, 
compared to our young sources and standards 
in the 1--3 Myr old Ophiuchus and Chamaeleon clouds, 
it appears that our NaI index 
may make the transition from the low value typical of giants and young cluster 
brown dwarfs to the higher values seen in field 
dwarfs at a younger age 
than the age at which the H band continuum transitions from the more
 triangular shape seen in younger objects to the flat-topped shape seen in 
older field dwarfs.
As a result, the Na index might be a more reliable indicator of cluster
 membership for the youngest clusters than the H-band 
continuum shape.
Based on Figure \ref{indices}, 
an NaI index of less than 1.1 (for spectral types later than M6) 
safely indicates that a source is 
low-gravity (either a young object or a giant).  
A spectral S/N of $\sim$80 is needed per resolution element 
(at R$\simeq$300) to determine if an object is 
low-gravity
based on a 100$\sigma$ determination of the NaI index.  
Changing the A$_V$ by 
which the spectra are dereddened by $\pm$1 magnitude changes the NaI 
indices by only 0.006; small uncertainties in A$_V$ would not change the 
classification of an object.  The wavelength of our NaI index lies on a prominent H$_2$O telluric absorption feature.  
To test the effect of atmospheric transmission on the calculation of our NaI 
index, we convolved our object and telluric standard star spectra with models 
of telluric atmospheric transmission \citep{atmo} of varying airmass (1--2) 
and precipitable water vapour content 
(pwv; 2--8~mm).  Differences in airmass of 0.15 and in pwv of 
10\% between the object and the standard star spectra resulted in only 
small ($\sim$1\%) changes in the calculated index.
All of our objects have NaI indices that are well separated 
(more than 0.1 in the index) from the field dwarf indices, providing further 
confirmation that our sources are indeed young, low-gravity objects.

\section{T$_{eff}$, Mass, and Age Estimates}
A motivation behind determining the spectral types of our sources is to derive an estimate of their effective temperatures. 
With effective temperatures and bolometric luminosities, 
we can place our objects on an H-R diagram, and use model 
isochrones \citep{chabrier00,dantona97} to estimate masses and ages.
\citet{allers06} obtained temperature and mass estimates
for our objects by assuming ages of 1~Myr and 3~Myr for  
sources in Ophiuchus and Chamaeleon II respectively, and assigning  
the parameters (T$_{eff}$, mass, and log~$g$) 
of the closest luminosity point on the appropriate \citet{baraffe03} 
isochrone to the observed luminosity.  In this section, we convert the 
spectral types to T$_{eff}$ and use these T$_{eff}$'s, together with the 
measured luminosities, to place the sources on an HR diagram for 
comparison with theoretical brown dwarf isochrones.

For the 6 sources in our sample, we estimate T$_{eff}$ from the 
spectral type of our objects using the SpT--T$_{eff}$ relationship of 
\citet{sluhman03}.  
For objects with spectral 
types from M6 to M9, we use the \citet{sluhman03} scale directly.  At a 
spectral type of L1, however, cha1305-7739 lies beyond the end of the  
\citet{sluhman03} temperature scale, so we subtract 
the difference in effective temperature between M9 and L1 field 
dwarfs \citep[196~K;][]{golimowski04} 
from the \citet{sluhman03} T$_{eff}$ for young M9 brown dwarfs 
(2400 K) and assign a temperature of 2200 K to cha1305-7739.
The T$_{eff}$'s of our sample based on their spectral 
types are listed in Table \ref{teff}.  The T$_{eff}$'s estimated from 
the objects' spectral types are higher  
than the \citet{allers06} T$_{eff}$'s, slightly so ($\lesssim$100 K) in 
two cases, but more 
significantly (200-500 K) for the remaining objects.

We have calculated the luminosities of our objects using the method described 
in \citet{allers06}, but using the values of A$_V$ in Table \ref{teff}.  
The positions of our Ophiuchus sources are within the boundary of the  
Upper Scorpius region, so we use a distance of 
145$\pm$20~pc \citep{preibisch06} to 
calculate the luminosities of oph1623-2338 and oph1622-2405, rather than the 
125~pc distance (appropriate for the Ophiuchus cloud core) 
adopted by the ``Cores to Disks'' Spitzer Legacy team and 
used by \citet{allers06}.  
The uncertainties in luminosity, including photometric, A$_V$, and distance 
uncertainties, total 0.10 and 0.13~dex for our sources in Chamaeleon II and 
Ophiuchus/Upper Sco respectively.  Our calculated luminosities (Table \ref{teff}) agree to within 
0.2 dex with the \citet{allers06} luminosities for all of the sources in our 
sample.  

\begin{figure}
\plotone{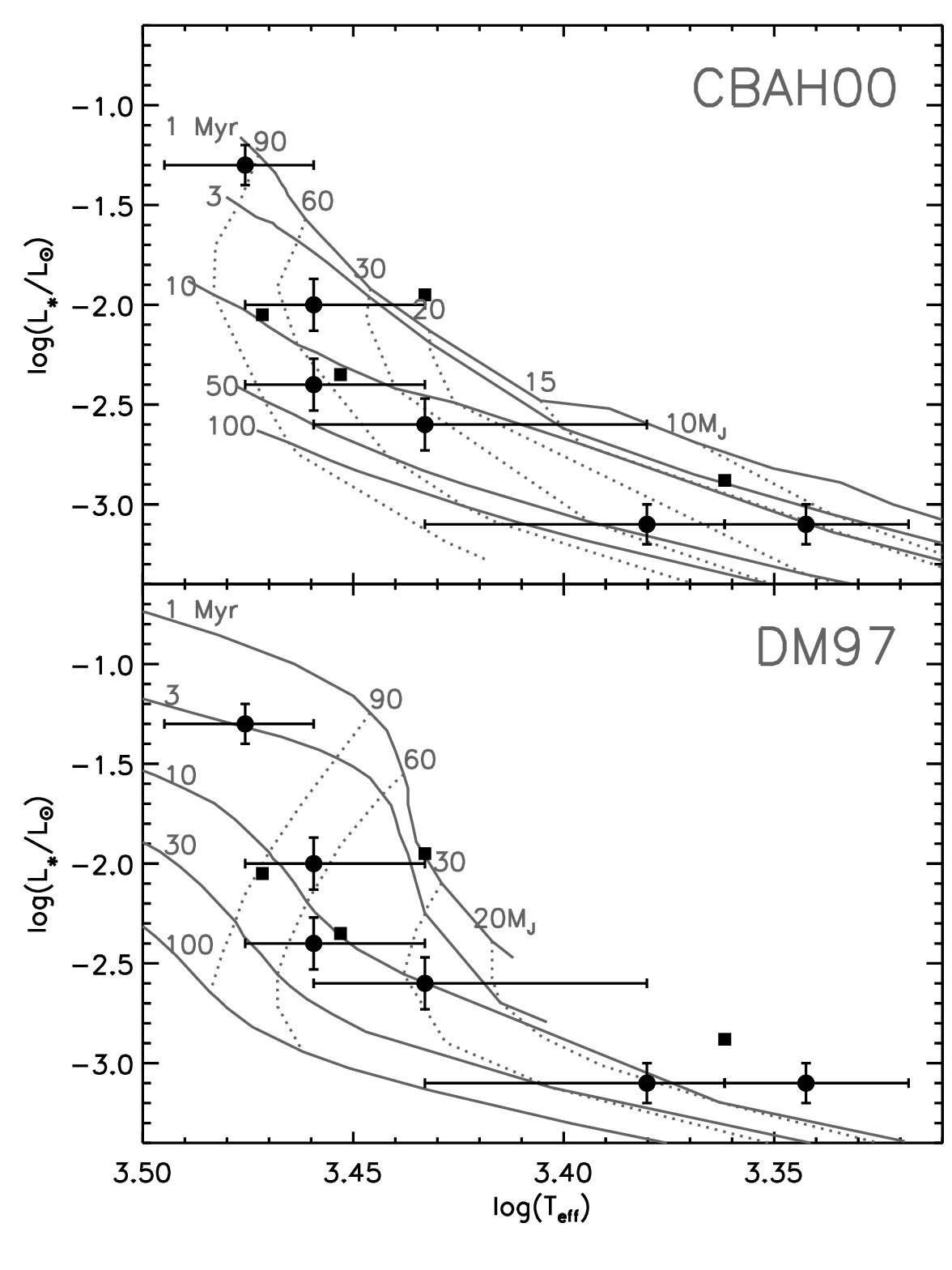}
\caption[H-R Diagram]
{\label{hr} H-R diagram for our sources.  The solid circles show
the positions of our sources with T$_{eff}$'s determined from the spectral
types of our objects and luminosities calculated using the
method described in \citet{allers06}.
Uncertainties in T$_{eff}$ are based of an uncertainty of
$\pm$~1~sub-type using the SpT-T$_{eff}$ relationships of \citet{sluhman03} and \citet{golimowski04}.
Uncertainties in luminosity are 0.10 dex for sources in Chamaeleon II and
0.13 dex for sources in Ophiuchus.  For comparison,
the filled squares
denote our young standards in Chamaeleon I, with values of luminosity and
T$_{eff}$ from \citet{luhman04c} and \citet{luhman04a}.
Top: The evolutionary models overlaid are the DUSTY model isochrones of \citet{chabrier00}.  Bottom:  The evolutionary models overlaid are the 1998 update of the \citet{dantona97} model isochrones for a deuterium abundance of 2$\times10^{-5}$.
}
\end{figure}

Figure \ref{hr} shows the placement of our objects on the H-R diagram.  
The error bars showing the uncertainties 
in T$_{eff}$ correspond to  
$\pm$~1 sub-type \citep{sluhman03, golimowski04} but assume that the 
SpT-T$_{eff}$ relation is correct. 
Table \ref{teff} shows the ages and masses of our sources inferred from 
both the DUSTY evolutionary models \citep[][hereafter CBAH00]{chabrier00} and the evolutionary models of 
\citet[][hereafter DM97]{dantona97} overlaid on the H-R diagram.  
In top panel of Figure \ref{hr}, 
the \citet{allers06} masses are roughly equivalent to moving 
horizontally from each source's position to the 3 Myr (for Chamaeleon II) 
and 1 Myr (for Ophiuchus) isochrones.
The derived ages and therefore the masses of our two brightest sources 
(oph1623-2338 and cha1258-7709), as determined from the CBAH00 model 
isochrones (5 and 1~Myr respectively), are in good 
agreement with the 1~Myr and 3~Myr ages assumed 
for Ophiuchus and Chamaeleon II in \citet{allers06}.  Using the DM97 model 
isochrones yields ages for oph1623-2338 and cha1258-7709 (7 and 3~Myr 
respectively) that are in good agreement with the CBAH00 model results, 
however, the masses estimated from DM97 are 1.6 times greater than those 
estimated from CBAH00 for these two sources.
The luminosity and temperature of our faintest source, cha1305-7739, do not 
fall within the DM97 model parameter space, but
the age and mass we derive using the CBAH00 isochrones 
(10~Myr, 15~M$_J$) agree to well within the uncertainties with the 
age of the Chamaeleon II cloud and the 10~M$_J$ mass from \citet{allers06}.  
The other three sources 
(cha1257-7701, oph1622-2405B and oph1622-2405A) have implied ages of 
20--30~Myr (CBAH isochrones) or 10--15~Myr (DM97 isochrones), and masses 
$\sim$2--4 times larger than those reported by \citet{allers06}.
Given that two of these sources 
(cha1257-7701 and oph1622-2405B) show mid-IR excess emission, 
presumably from 
a circumstellar disk, and that typical lifetimes for T~Tauri disks are 
$\lesssim$6~Myr \citep{haisch01}, it is unlikely that these sources are 
20 and 30~Myr old if brown dwarf disks have similar lifetimes as T~Tauri disks.
Our near-IR spectra of these sources also 
show that they must be young (based on their H-band continuum shape and NaI 
indices).  The discrepancy between the \citet{allers06} and our 
derived T$_{eff}$'s, masses and ages from the same group of model isochrones 
could be a result of 
1) our choice of SpT--T$_{eff}$ relation, 2) an underestimation of the source luminosities, or 3) an older population of brown dwarfs associated with 
the Chamaeleon II and Ophiuchus regions that still show signatures of youth.  
An additional source of uncertainty in the derived masses and ages of our objects lies in the evolutionary models themselves.
Further discussion of each of these possibilities follows.

\subsection{Uncertainties in the SpT--T$_{eff}$ Relation}

A possible source of the inconsistency in the effective temperatures and masses from 
\citet{allers06} and those in Figure \ref{hr} lies in the sensitivity of the 
derived parameters to the choice of SpT--T$_{eff}$ relation. 
The best available relation is a compromise between the SpT-T$_{eff}$ relation for 
field dwarfs and that for giants \citep{luhman99, sluhman03}.  The intermediate 
position was determined by requiring all four components of the TTau system to lie on the 
same isochrone in the models of \citet{baraffe98}.  The sources lie along the 1 Myr 
isochrone, consistent with ages of other Taurus YSOs.  Interestingly, the 1 Myr 
isochrone is parallel to the 3 Myr isochrone at both higher and lower temperatures, but, 
just in the temperature range relevant for TTau Bb ($\sim$2800~K), the 1 Myr isochrone jumps closer 
to the 3 Myr line \citep{baraffe98}.  If this anomaly were not present, the coevality 
requirement would have forced the later spectral types 
to lower temperatures.  
This would bring the present results into better agreement with the results of 
\citet{allers06}.  Uncertainties in the SpT--T$_{eff}$ relation, however, cannot explain 
the low luminosity of some of our sources relative to young standards
of similar spectral type (as seen in Figure \ref{hr}). 

\begin{deluxetable*}{l|r|r|rrcrrrrr}
\tablecolumns{11}
\tabletypesize{\footnotesize}
\tablecaption{Derived Physical Properties of our Sample}
\tablewidth{0pt}
\tablehead{
\multicolumn{7}{c}{} &
\multicolumn{2}{c}{CBAH00} &
\multicolumn{2}{c}{DM97}  \\
\colhead{Source} &
\colhead{$\alpha$(J2000)}             &
\colhead{$\delta$(J2000)}             &
\colhead{SpT} &
\colhead{A$_V$} &
\colhead{L$_{\ast}$} &
\colhead{T$_{eff}$\tablenotemark{a}} &
\colhead{M$_{\ast}$\tablenotemark{b}}  &
\colhead{Age\tablenotemark{b}}  &
\colhead{M$_{\ast}$\tablenotemark{c}}  &
\colhead{Age\tablenotemark{c}}  \\
\colhead{} &
\colhead{} &
\colhead{} &
\colhead{} &
\colhead{mag} &
\colhead{log $L_{\ast}/L_{\odot}$} &
\colhead{K} &
\colhead{M$_{\rm J}$} &
\colhead{Myr} &
\colhead{M$_{\rm J}$} &
\colhead{Myr} 
}
\startdata
cha1257-7701 & 12 57 58.7& -77 01 19.5&M9&4.0 &-3.1&2400&30&30&20 &15\\
cha1258-7709 & 12 58 06.7& -77 09 09.5&M6&10.4&-1.3&2990&90&1 &140&3\\
cha1305-7739 & 13 05 40.8& -77 39 58.2&L1&3.0 &-3.1&2200&15&10&-- &--\\
oph1622-2405B& 16 22 25.2& -24 05 15.6&M8&1.0 &-2.6&2710&35&20&30 &10\\
oph1622-2405A& 16 22 25.2& -24 05 13.7&M7&0.0 &-2.4&2880&65&20&55 &15\\
oph1623-2338 & 16 23 05.8& -23 38 17.8&M7&8.2 &-2.0&2880&40&5 &65 &7\\
\enddata
\label{teff}
\tablenotetext{a}{From the SpT-T$_{eff}$ relationships of \citet{sluhman03} and \citet{golimowski04}}
\tablenotetext{b}{Estimated from the DUSTY model isochrones of \citet{chabrier00}}
\tablenotetext{c}{Estimated from the 1998 update of the 
\citet{dantona97} isochrones.  Cha1305-7739 does not fall in the parameter space of the \citet{dantona97} models.}
\end{deluxetable*}
  
\subsection{Underestimating the Luminosity}

\citet{allers06} rigorously tested
their method of calculating L$_{\ast}$ using field M6-L6 dwarfs with known distances and hence accurate luminosities \citep{golimowski04} and found 
uncertainties well below 5\%.  It might be possible that the luminosities of our 
3 discrepant objects are underestimated either because the objects are  
more distant than the cloud 
\citep[145 pc for Ophiuchus and 178 pc for Chamaeleon II;][]{preibisch06,whittet97} or because
they are detected in scattered light.  
In the case of cha1257-7701, our derived extinction (A$_V$=4.0), 
is greater than the total 
cloud extinction in the vicinity of cha1257-7701 \citep[A$_V \sim$2;][]{cambresy99}, 
so it 
is possible that the source lies slightly behind the Chamaeleon II cloud.  
In order for the source luminosity to lie on the 3 Myr CBAH00 isochrone 
for a T$_{eff}$ of 2400~K, however, 
cha1257-7701 would have to lie $\sim$90~pc behind the cloud, which seems 
unlikely 
given that the Chamaeleon II cloud is only $\sim$5~pc wide, and 
presumably of similar depth.  Oph1622-2405A and B lie 
close to the boundary defining the Upper Scorpius association 
\citep[d=145$\pm$20~pc;][]{preibisch06}.  
If Oph1622-2405A and B lie behind the Ophiuchus cloud and 
are members of Upper Scorpius, using the maximum distance for sources in 
Upper Sco, 165~pc (rather than 125~pc as used in 
\citet{allers06}) to calculate the luminosities would place them close to the 
10 Myr isochrone in Figure \ref{hr}, in reasonable agreement with the age of 
Upper Scorpius \citep[$\sim$5~Myr][]{preibisch02}. 
Oph1622-2405A and B show no sign of foreground reddening, even though the A$_V$ of the 
Ophiuchus cloud is $\sim$4.5 magnitudes \citep{ridge06} 
at the position of oph1622-2405A and B,
 which makes it unlikely that oph1622-2405A and B lie behind the 
Ophiuchus/Upper~Sco cloud material.
Detection in scattered light is another possible explanation for the low luminosity of the
three discrepant sources.
However, our full 
0.8--24.0~$\mu$m SED's show no indication of wavelength-dependent 
Rayleigh scattering (i.e. their colors are in good agreement with models 
for a brown dwarf with a passively heated disk).  The two sources showing mid-IR excess 
also agree with the constant L$_{\rm excess}$/L$_{\ast}$ found by \citet{allers06}, 
which would be unlikely if they are detected in scattered light.  Optical 
spectra of oph1622-2405 and cha1257-7701 \citep{binary,jayawardhana06}, 
which would be more 
more sensitive to the effects of Rayleigh scattering than our near-IR spectra,
also show no evidence for detection in scattered light.

\subsection{An Older Cluster Population}

Another possibility is that these three sources could indeed be older than the 
main population in their associated clusters.  An older population has been 
suggested as a possible explanation for bifurcation on the H--R diagram 
of sources in the 
Orion Nebula Cluster \citep{palla05,slesnick04}. 
Since two of our three discrepant 
objects are 
known to harbor circumstellar disks, this might seem unlikely given the 
relatively short 
lifetime of T~Tauri disks \citep[$\sim$6~Myr][]{haisch01}.
If the disk is depleted mainly by 
accretion, however, it may be possible for brown dwarf disks to be longer lived than 
T~Tauri disks.  Mass accretion rates (based 
on H$\alpha$ emission) depend strongly on the mass of the central source, 
with $\dot{M}$ $\propto$ M$_{\ast}^{1.8-2.1}$ for masses 
from 150~M$_{\rm J}$ down to roughly
25 M$_{\rm J}$ \citep{muzerolle05,natta06}.  If the disk is depleted mainly via 
accretion onto the central source, the $\dot{M}$--M$_{\ast}$ 
relation indicates that the 
timescale for depleting the disk is inversely proportional to the stellar mass, 
or T~$\propto$~M$_{\ast}^{-0.8-1.1}$.  Thus, a 50 M$_{\rm J}$ object could have 
a disk for 2.1 times longer than a 100~M$_{\rm J}$ object.  
The \citet{haisch01} 
lifetimes were determined by observing disk fractions (as measured by 
L$^{\prime}$ excess) towards clusters with ages from 0.3--30~Myr.  The younger 
clusters in their sample have lower mass completion limits than 
the older clusters in their sample both because sources of a given mass are 
brighter at younger ages and because the older clusters in their survey are 
3--4 times more distant than their young clusters.  
Thus, the measured disk fractions for the 
young clusters in their survey may be artificially higher than 
their older clusters if lower-mass 
objects do indeed have longer disk lifetimes.  
Oph1622-2405B is 
$\sim$4 times less massive than the lowest mass sources in \citet{haisch01} 
and $\sim$30 
times less massive than the 1 M$_{\odot}$ completeness 
limit for their 5~Myr old sources 
in NGC~2362.  Based on the \citet{haisch01} assumptions, along with the 
longer disk lifetimes for young brown dwarfs implied by their mass accretion 
rates, it is plausible that a 30--40~M$_{\rm J}$ object could have a disk for 
as long as 30~Myr.  As mentioned in \S3.2.3 and 3.2.4, oph1622-2405A and B have slightly 
stronger NaI features than the young standards to which they are compared, 
which might hint at a higher gravity and somewhat older age, in agreement 
with the $\sim$5~Myr age independently determined for oph1622-2405 
by \citet{binary} and \citet{close07}.  
Our spectrum of cha1257-7701 has too low of S/N for comparison of gravity sensitive atomic lines.

\subsection{Uncertainties in the Evolutionary Models}

Table \ref{teff} shows the masses and ages of our sources derived using two 
different sets of evolutionary models, CBAH00 and DM97.  Visual inspection 
of Figure \ref{hr} reveals that the slopes and positions of the two sets of 
isochrones and mass tracks differ substantially.  Not surprisingly, the 
masses and ages derived from the two sets of isochrones vary by as much as a 
factor of $\sim$1.6 in mass and a factor of $\sim$3 in age.  
The descrepancy in 
mass is highest (a factor of 1.6) for the brightest source in our sample, 
cha1258-7709, whereas the masses of the lower luminosity sources fall 
within the error bars.
The spread in source ages is smaller when derived from the DM97 models 
(3--15~Myr) than the spread in ages determined using the CBAH00 
models (1--30~Myr).  The differences in the masses and ages of our sample as determined from the DM97 and CBAH00 models 
are not surprising given that evolutionary models, in general, are unable to 
reproduce the dynamically determined masses of pre-main-sequence stars to 
consistently better than $\sim$30\% \citep{hillenbrand04}.  For both sets of 
evolutionary models the isochrones and mass tracks become almost parallel at 
the lowest temperatures, implying that their power to distinguish age 
differences from mass differences is diminished.

\section{Summary}

We have presented follow-up near-IR spectroscopy of 5 candidate objects with 
circumstellar disks reported in \citet{allers06} as well the spectrum 
of a probable companion to oph1622-2405B.  The spectral features 
confirm the young age of these objects.  
We are able to use our R$\simeq$300 0.8-2.4 $\mu$m spectra to 
establish the spectral types of 
our objects by comparison to young brown dwarfs in Chamaeleon I with spectral 
types known from optical spectra \citep{luhman04c, luhman04a} as well as to  
two young field dwarfs, 2MASS0141--4633 and G196--3B \citep{kirkpatrick06,kirkpatrick01,rebolo98}.  The results are accurate to $\pm$1 subtype.
Our spectrum of cha1305-7739 indicates that it has a spectral type of $\sim$L1, 
making it among the latest type young brown dwarfs with 
circumstellar disks discovered to date.  

Comparing spectra of young brown dwarfs, field dwarfs and giants, we 
found an H$_2$O index capable of determining spectral type to $\pm$1 sub-type 
independent of gravity.  We also created an index based on the 1.14~$\mu$m 
NaI feature that is sensitive to gravity, but only weakly dependent on 
spectral type for field dwarfs.  Our NaI index appears to be a more sensitive 
gravity indicator than the triangular H-band continuum shape, and can 
distinguish between young sources in clusters with $\tau\lesssim$5~Myr and 
young members of the field population with ages of tens of Myr.

Placing our 
objects on the H-R diagram using the luminosities from \citet{allers06} 
and T$_{eff}$'s from our spectral types using the SpT-T$_{eff}$ relationship 
of \citet{sluhman03} gives masses and ages consistent with the \citet{allers06} 
results for half of our sources but yields older ages and masses up to 5.5 times 
larger for three of our sources.  
The cause of the discrepancy could lie in the SpT-T$_{eff}$ relation, an underestimation of 
the distance to the sources, 
the detection of the sources in scattered light, or the existence 
of a bona-fide older population of brown dwarfs with circumstellar disks.
The large differences in derived
masses (from the same set of model isochrones) using two independent but reasonable methods implies that we need to be quite circumspect in quoting masses for young brown dwarfs.  Refinements to the
SpT-T$_{eff}$ relation and/or spectroscopic techniques for direct determination of surface gravity may reduce the age and mass problem significantly.
Uncertainties in the evolutionary models themselves can be quite large.  Using two separate sets of model isochrones \citep{chabrier00,dantona97} to derive the physical properties of our sample yields differences as large as a factor of 1.6 in mass and a factor of 3 in age.

\acknowledgments
We are grateful to J.~D. Kirkpatrick for 
providing the spectrum of 2MASS0141-4633.
KNA and MCL acknowledge support for this work from
NSF grant AST-0407441.  MCL also acknowledges support from an Alfred P.
Sloan Research Fellowship.
KL was supported by grant NAG5-11627 from the NASA Long-Term Space
Astrophysics program.

\end{document}